\documentclass[useAMS,usenatbib]{mn2e}

\usepackage{hyperref}
\usepackage{graphicx}
\usepackage{amssymb}
\usepackage{amsmath}
\usepackage[english]{babel}
\usepackage[T1]{fontenc}
\usepackage{aecompl}
\usepackage{txfonts}

\def\zphot{$z_{\text{phot}}$}
\def\zspec{$z_{\text{spec}}$}

\title[\zphot\, in KiDS-ESO-DR2]{Machine Learning based photometric redshifts for the KiDS ESO DR2 galaxies}

\author[S. Cavuoti et al.]{S.~Cavuoti$^{1}$\thanks{E-mail:
stefano.cavuoti@gmail.com}, M.~Brescia$^{1}$, C.~Tortora$^{1}$, G.~Longo$^{2}$, N.~R.~Napolitano$^{1}$, M.~Radovich$^{3}$, \and F.~La Barbera$^{1}$, M.~Capaccioli$^{4}$, J.~T.A.~de Jong$^{5}$, F.~Getman$^{1}$, A.~Grado$^{1}$, M.~Paolillo$^{2}$.\\
$^{1}$Astronomical Observatory of Capodimonte - INAF, via Moiariello 16, I-80131, Napoli, Italy\\
$^{2}$Department of Physics, University Federico II, via Cinthia 6, 80126 Napoli, Italy\\
$^{3}$Astronomical Observatory of Padua, vicolo dell'Osservatorio 5, I-35122 Padova, Italy\\
$^{4}$VSTCen - INAF, via Moiariello 16, I-80131, Napoli, Italy\\
$^{5}$Leiden Observatory, Leiden University, P.O. Box 9513, 2300 RA Leiden, The Netherlands}

\date{Accepted 2015 July 03; Received 2015 July 02; in original form 2015 May 05}
\pagerange{\pageref{firstpage}--\pageref{lastpage}} \pubyear{2015}

\begin{document}
\phantomsection
\label{firstpage}
\maketitle

\begin{abstract}
We estimated photometric redshifts (\zphot) for more than $1.1$ million galaxies of the ESO Public Kilo-Degree Survey (KiDS) Data Release 2. KiDS is an optical wide-field imaging survey carried out with the VLT Survey Telescope (VST) and the OmegaCAM camera, which aims at tackling open questions in cosmology and  galaxy evolution, such  as the origin of dark energy and the channel of galaxy mass growth. We present a catalogue of photometric redshifts obtained using the Multi Layer Perceptron with Quasi Newton Algorithm (MLPQNA) model, provided within the framework of the DAta Mining and Exploration Web Application REsource (DAMEWARE). These photometric redshifts are based on a spectroscopic knowledge base which was obtained by merging spectroscopic datasets from GAMA (Galaxy And Mass Assembly) data release 2 and SDSS-III data release 9. The overall $1\sigma$ uncertainty on $\Delta z = (z_{spec} - z_{phot}) / (1+ z_{spec})$ is $\sim 0.03$, with a very small average bias of $\sim 0.001$, a NMAD of $\sim 0.02$ and a fraction of catastrophic outliers ($\left| \Delta z \right| > 0.15$) of $\sim 0.4 \%$.
\end{abstract}


\section{Introduction}

Photometric redshifts (\zphot) derived from multi-band digital surveys are crucial to a variety of cosmological applications \citep{scranton2005,myers2006,hennawi2006,giannantonio2008}. In the last years a plethora of methods has been developed to estimate \zphot\, (cf. \citealt{hildebrandt2010}), but the advent of a new generation of photometric surveys (to quote just a few, Pan-STARRS: \citealt{kaiser2004}; Euclid: \citealt{laureijs2011};
KiDS\footnote{\url{http://kids.strw.leidenuniv.nl/}}: \citealt{dejong2013}) demands for higher accuracy \citep{brescia2014}.

The evaluation of photometric redshifts requires the mapping of the photometric space into the spectroscopic redshift space. All methods, one way or the other, require the use of a Knowledge Base (KB) consisting in a set of templates, and differ mainly in the following aspects: \textit{(i)} the way in which the KB is constructed (spectroscopic redshifts or, rather, empirically or theoretically derived spectral energy distributions or SEDs), and \textit{(ii)} the adopted interpolation/fitting algorithm. Methods based on the interpolation of a spectroscopic KB are usually labeled as empirical.

Many different implementations of empirical methods exist and we shall recall just a few: \textit{i)} polynomial fitting \citep{connolly95}; \textit{ii)} nearest neighbors \citep{csabai03}; \textit{iii)} neural networks (\citealt{dabrusco07,yeche10} and references therein); \textit{iv)} support vector machines \citep{wadadekar05}; \textit{v)} regression trees \citep{carliles2010}; \textit{vi)} gaussian processes \citep{way06,bonfield10}, and \textit{vii)} diffusion maps \citep{freeman09}.

In this paper we discuss the derivation of photometric redshifts for the galaxies in the Kilo-Degree Survey (KiDS) data release 2 \citep{dejong2015}. KiDS is an ESO public survey, based on the VLT Survey Telescope \citep{cap2011} with the OmegaCAM camera \citep{kuijken2011}, that will image $1500$ square degrees in four filters (u, g, r, i), in single epochs per filter. The high spatial resolution of VST ($0.2\arcsec / pixel$), the photometric depth and area covered make it a front-edge tool for weak gravitational lensing and galaxy evolution studies. The measurement of unbiased and high-quality \zphot\, is a crucial step to pursue many of the scientific goals which motivated the KiDS survey \citep{dejong2015}.

We present the \zphot\, for a sample of $\sim 1.1$ million galaxies. These redshifts were derived with the Multi Layer Perceptron with Quasi Newton Algorithm (MLPQNA) method, described in detail elsewhere \citep{brescia1, brescia2}, hence we refer to those articles for all the mathematical and technical details. Recently, this method has been also used to derive a catalogue of \zphot\, for the entire SDSS-DR9 \citep{brescia2014}. In the PHAT1 contest \citep{hildebrandt2010}, which blindly compared most existing methods to estimate \zphot\, on a very limited KB ($\sim500$ objects only), the MLPQNA method proved to be one of the best empirical methods to date \citep{cavuoti0}. It is however worth noticing that in the PHAT1 contest, MLPQNA did not perform as well as many SED fitting methods, due to the very limited base of knowledge available. This situation reverses when significantly larger KB's properly sampling the photometric parameter space become available \citep{brescia2,brescia2014}.

The MLPQNA model is publicly available in the DAta Mining \& Exploration Web Application REsource infrastructure (DAMEWARE; \citealt{brescia3}) and has also been implemented in the PhotoRaptor service package \citep{cavuoti2015}.

The paper is organized as follows. In Sect.~\ref{thedata} we present the dataset, while in Sect.~\ref{experiments} the experiments and related outcome are described and discussed. In Sect.~\ref{catalog} we give a description of the resulting photometric redshift catalogue. Finally, in Sect.~\ref{conclusions} we draw our conclusions and
future prospects.

\section{The data}\label{thedata}

The sample of galaxies for which we provide \zphot\, was extracted from the second data release of the ESO Public Kilo-Degree Survey (KiDS-ESO-DR2). A detailed description of all the steps followed to extract the catalogues is given in \cite{dejong2015}. KiDS is a wide-area optical imaging survey in the four filters ($u$, $g$, $r$, $i$), carried out with the VLT Survey Telescope (VST) and the OmegaCAM camera. The KiDS observation strategy consists of a standard diagonal dithering pattern ($5$ dithers in $g$, $r$, $i$ and $4$ in $u$-band) in order to minimize the effect of the inter-CCD gaps in the OmegaCAM science array. Therefore the final footprint of each single tile is slightly larger than the nominal $1$ square degree \citep{dejong2015}.

The data processing procedure used is based on the Astro-WISE (AW) optical pipeline (\citealt{McFarland+13_AstroWise}). After the first basic data reduction steps (such as cross-talk, de-biasing and overscan correction, flat-fielding, illumination correction, de-fringing, pixel masking, satellite-track removal and background subtraction), the pipeline performs photometric and astrometric calibrations.

Source extraction is based on a task provided in the AW environment, KiDS-CAT \citep{dejong2015}, based on algorithms developed for the software 2DPHOT \citep{LaBarbera_08_2DPHOT}. KiDS-CAT automatically performs a seeing assessment of the image, using best-quality stars in the image, and subsequently optimizes the configuration files of SExtractor \citep{Bertin_Arnouts96_SEx} to perform the source extraction in the individual bands. In this  process, besides  the photometric flag  provided by  SExtractor, detected sources  are also flagged  according  to  their  proximity to  star  spikes  and  haloes (IMAFLAGS\_ISO flag), which  are identified in the KiDS  images through a dedicated masking  procedure (\citealt{dejong2015}, see also  Huang et al. in preparation).

In order to derive our photometric redshifts, we used the multi-band source catalogues, which rely on the double-image mode of SExtractor. These catalogues are based on source detection in the r-band images. While magnitudes are measured in all filters, the Star-Galaxy separation, as well as the source positional and shape parameters, are based on the r-band data only. The choice of the r-band as a reference is motivated by the fact that it is observed under the best seeing conditions ($\sim0.7\arcsec$ seeing FWHM, on average), and therefore it typically has the best image quality, thus providing the most reliable source positions and shapes. Aperture photometry in the four bands within several aperture radii, together with MAG\_AUTO, shape parameters and flags, are available from SExtractor and KiDS-CAT. In the final catalogue, in order to maximize the sample with \zphot\, estimates available, we have retained $\sim 10^7$ sources with r-band SExtractor flag $FLAGS\_r < 4$ and rejected $\sim 2\times10^5$ objects having close and bright companion sources, affected by bad pixels or originally blended with other objects (see~\citealt{Bertin_Arnouts96_SEx} for a detailed description of extraction flags). The limiting magnitudes of KiDS catalogues\footnote{We use the MAGAP\_4 and MAGAP\_6 magnitudes, measured within circular apertures of 4\arcsec and 6\arcsec of diameter, respectively. These magnitudes are provided within the produced \zphot\, catalogue.} at the $1\sigma$ level are:

\begin{itemize}
 \item  MAGAP\_4\_u $=25.17 $
 \item  MAGAP\_6\_u $=24.74 $
 \item  MAGAP\_4\_g $=26.03 $
 \item  MAGAP\_6\_g $=25.61 $
 \item  MAGAP\_4\_r $=25.89 $
 \item  MAGAP\_6\_r $=25.44 $
 \item  MAGAP\_4\_i $=24.53 $
 \item  MAGAP\_6\_i $=24.06 $
\end{itemize}

KiDS DR2 contains $148$ tiles observed in all filters during the first two years of operations. From the original catalogue of $\sim 18$ millions of sources, the Star/Galaxy separation leaves $\sim 10$ million galaxies, of which $\sim 6$  million have null IMAFLAGS\_ISO in all the filters, i.e. they are observed in unmasked regions. Out of these, we succeeded in estimating \zphot\, for $1,142,992$ sources (see Sec.~\ref{catalog} for details).

In order to build the needed spectroscopic knowledge base, the KiDS galaxy sample was matched with two independent spectroscopic surveys: GAMA (Galaxy And Mass Assembly) and SDSS (Sloan Digital Sky Survey). The final spectroscopic sample was obtained by merging data from GAMA data release 2 ($112k$ new redshifts in the first three years, \citealt{Driver+11_GAMA}, Liske et al. in prep.) and SDSS-III data release 9 \citep{Ahn+12_SDSS_DR9,ahn2014,bolton2012,chen2012}. The redshift distribution of the mixed catalogue is shown in Fig.~\ref{histotraintestspec}.

GAMA observes galaxies out to $z=0.5$ and $r < 19.8$ (r-band petrosian magnitude), by reaching a spectroscopic completeness of $98\%$ for the main survey targets. It provides also information about the quality of the redshift determination by using the probabilistically defined \textit{normalised redshift quality scale} nQ. The redshifts with $nQ > 2$ are the most reliable \citep{Driver+11_GAMA,hopkins2013}.

For the SDSS-III we used the low $z$ (LOWZ) and constant mass (CMASS) samples of the Baryon Oscillation Sky Survey (BOSS). The BOSS project aims to obtain spectra (redshifts) for $1.5$ million luminous galaxies up to $z \sim 0.7$. The LOWZ sample consists of galaxies with $0.15<z<0.4$ with colors similar to those of luminous red galaxies (LRGs) at $z \gtrsim  0.4$. Objects were selected by applying suitable cuts on magnitudes and colors with the aim of extending the SDSS LRG sample towards fainter magnitudes/higher redshifts (see e.g.~\citealt{Ahn+12_SDSS_DR9, bolton2012}).

The CMASS sample contains three times more galaxies than the LOWZ sample, and it was designed to select galaxies in the range $0.4 < z < 0.8$. The rest-frame color distribution of the CMASS sample is significantly broader than that of the LRG one, thus CMASS contains a nearly complete sample of massive galaxies down to $\log  M_{\star}/ M_\odot \sim  11.2$. The faintest galaxies are at $r = 19.5$ in the LOWZ sample and $i = 19.9$ in the CMASS one.

Our spectroscopic sample is therefore dominated by galaxies from  GAMA ($46,603$ vs. $1,618$ from SDSS) at low-z ($z  \lesssim  0.4$),  while SDSS  galaxies  dominate  the  higher redshift regime (out to $z \sim 0.7$), with $r < 22$.

\section{Experiments and discussion}\label{experiments}
Dealing with machine learning supervised methods, it is common practice to select and use the available KB to build a minimum of three disjoint data subsets: (\textit{i}) a data set to train the method looking for the correlation hidden in the photometric information among the input features necessary to perform the regression (known as training set); (\textit{ii}) a validation set to be used to check and verify the training performance against a loss of generalization capabilities (a phenomenon also known as \textit{overfitting}); (\textit{iii}) finally, a test set needed to blindly evaluate the overall performances of the model with data samples never submitted to the model before.

In this work, the validation process was embedded into the training phase, by applying the standard leave-one-out k-fold cross validation mechanism \citep{geisser1975}. We would like to stress that none of the objects included in the training (and validation) sample were included in the test sample and only the test data were used to generate the statistics and scatter plots.

\begin{figure*}
   \centering
   \includegraphics[width=\textwidth]{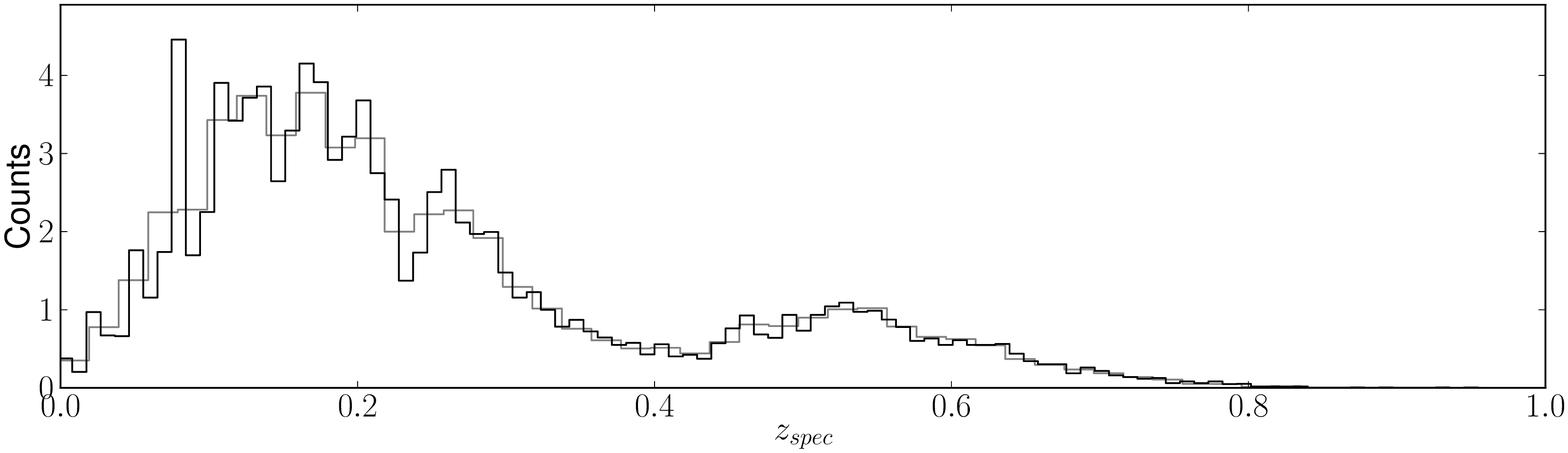}
   \caption{Spectroscopic redshift distribution of objects included in the training set (black line) and test set (gray line) normalized to the splitting rate.}\label{histotraintestspec}
\end{figure*}

\begin{figure*}
   \centering
   \includegraphics[width=\textwidth]{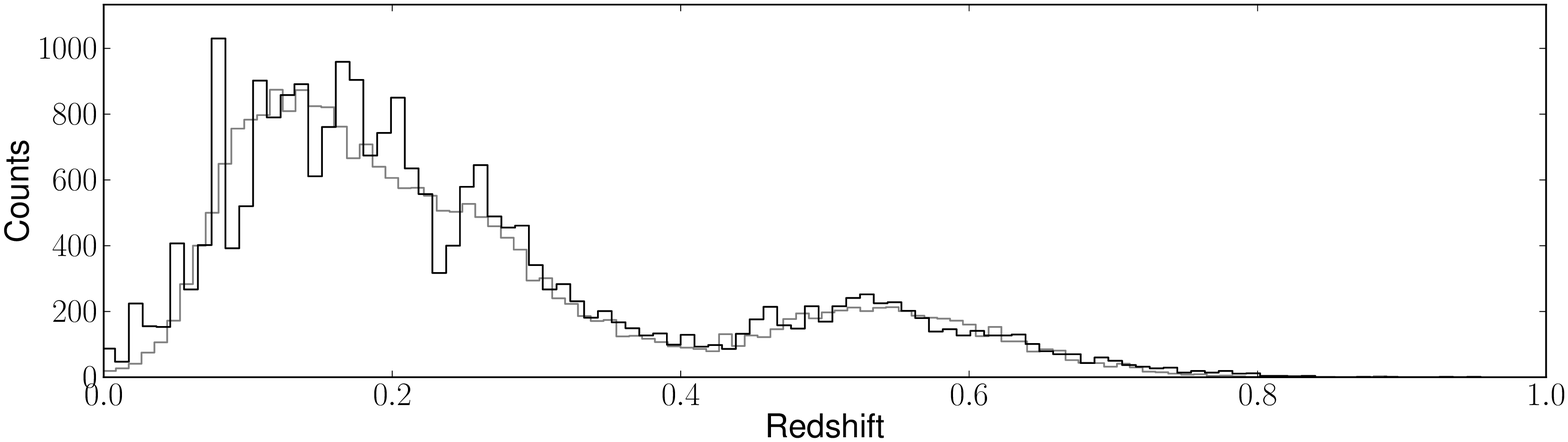}
   \caption{Redshift distribution of objects included in the blind test set, spectroscopic (black line) and photometric (gray line).}\label{histospec}
\end{figure*}

We created training and test samples with relative sizes of $60\%$ ($36,222$ objects) and $40\%$ ($24,150$ objects) by randomly drawing without replacement from the KB. The histogram in Fig.~\ref{histotraintestspec} shows the distribution of the KB as a function of the $z_{spec}$ in both the training and test sets, while in Fig.~\ref{histospec} the distribution of $z_{spec}$ and $z_{phot}$ in the blind test set is shown. As it can be seen, the three distributions are in almost perfect agreement.

The results were evaluated using a standard set of statistical indicators applied to the quantity $\Delta z = \frac{z_{spec} - z_{phot}}{1+z_{spec}}$:
\begin{itemize}
\item the bias, defined as the mean value of the residuals $\Delta z$;
\item  the standard deviation ($\sigma$) of the residuals;
\item the normalized median absolute deviation or $NMAD$ of the residuals, defined as $NMAD(\Delta z) = 1.48 \times Median\left( \left| \Delta z \right| \right )$.
\end{itemize}

As input photometric parameters (or features) we used the MAGAP\_4 and MAGAP\_6 aperture magnitudes ($u$, $g$, $r$, $i$), a choice which was based on our past experience, since almost always this combination lead to the best performances \citep{brescia2,brescia2014}. However, it needs to be emphasized that an improvement in the performances of a machine learning method can be expected from an exhaustive exploration of the parameter space through feature selection (cf. \citealt{polsterer2014}). This approach, however, is usually too much demanding in terms of computing time.

\begin{figure}
   \centering
   \includegraphics[width=9cm]{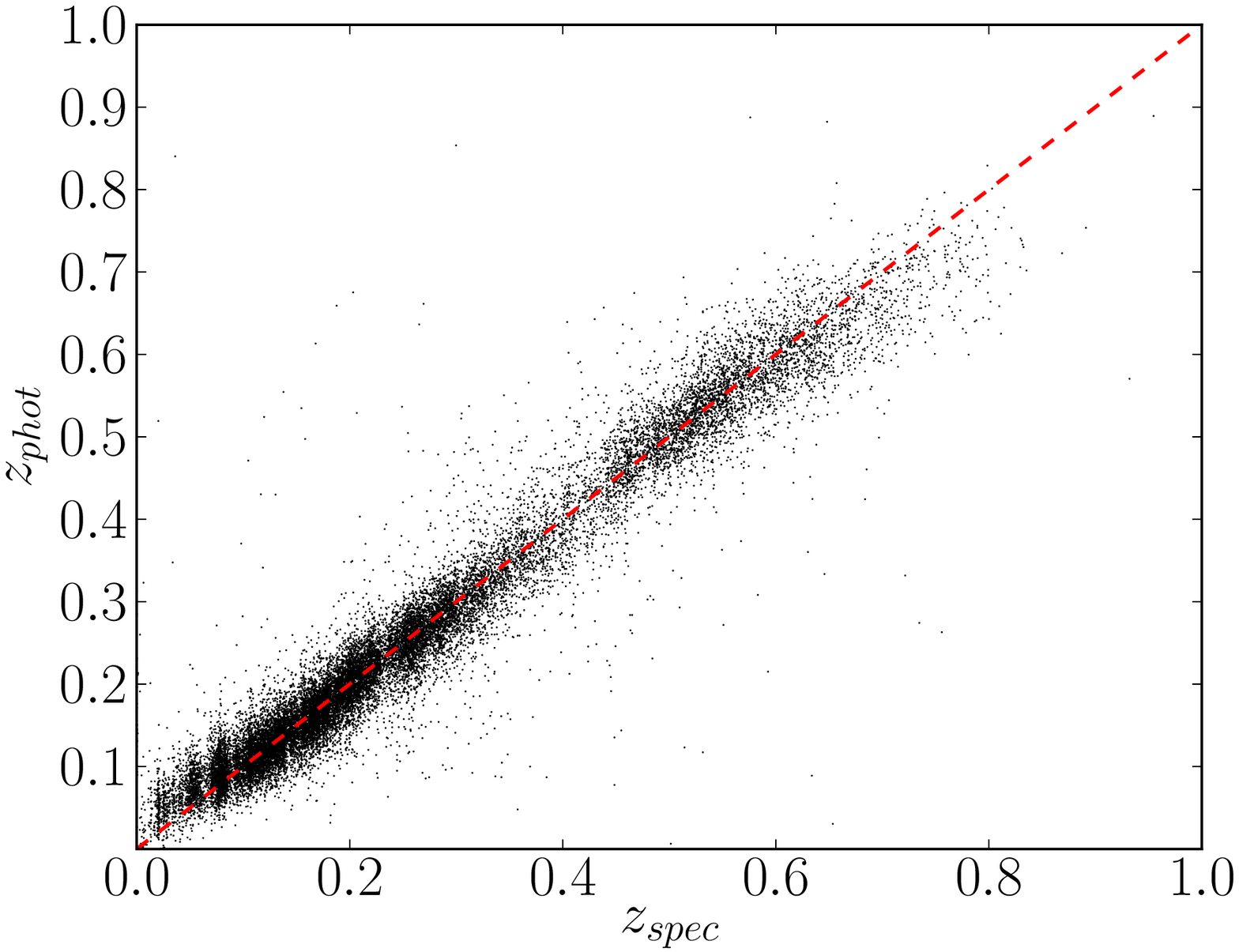}
   \includegraphics[width=9cm]{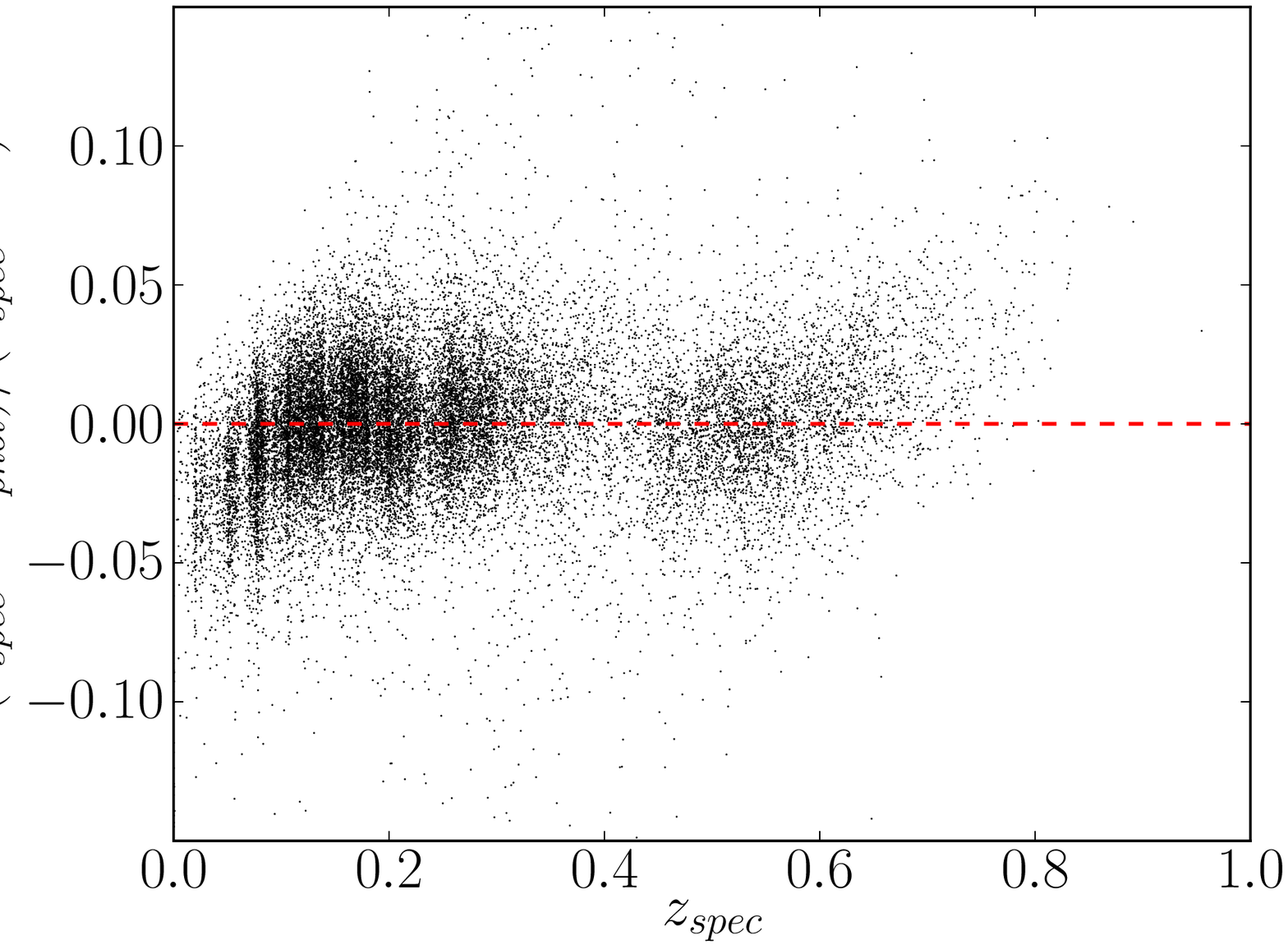}
   \caption{Upper panel: spectroscopic versus photometric redshifts for objects of the blind test set. Lower panel: spectroscopic redshift versus (\zspec-\zphot)/(1+\zspec) for the same objects.}\label{scattertest}
   \end{figure}

\begin{figure}
  \centering
  \includegraphics[width=9cm]{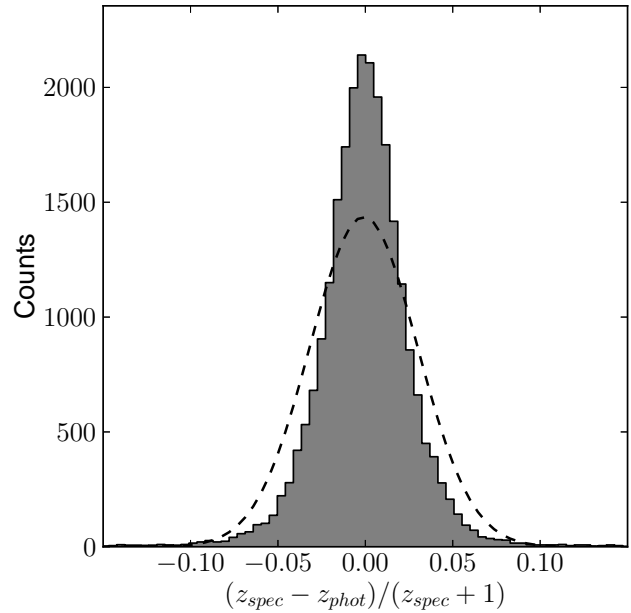}
  \caption{Histogram of (\zspec - \zphot)/(1+ \zspec) for objects of the blind test set. The dashed line represents the gaussian fit to the distribution.}\label{dzgauss}
  \end{figure}

MLPQNA \zphot\, are in excellent agreement with $z_{spec}$, as we show in Figs.~\ref{scattertest} and~\ref{dzgauss}, where the results of the
experiment are summarized. The upper panel of Fig.~\ref{scattertest} shows the predicted photometric redshift estimates versus the spectroscopic redshift values for all objects in the blind test set.
In the lower panel of  Fig.~\ref{scattertest} the $z_{spec}$ is plotted vs. the residuals $\Delta z$.
The underpopulated redshift bins, visible in Fig.~\ref{scattertest}, reflect the distribution of the spectroscopic sample which is less populated at redshift $\sim 0.22$ and $\sim 0.42$ (see Fig.~\ref{histotraintestspec} and Fig.~\ref{histospec}).

In Fig.~\ref{dzgauss} we show the distribution of residuals which has a kurtosis of $1.8$ and a skewness of $7.07 \times 10^{-16}$, i.e. a \textit{leptokurtic} and symmetric distribution, as already found in the SDSS-DR9 case by applying the same method \citep{brescia2014}.
In other words, the distribution reveals an over-density of objects in its central region (i.e. objects with a small error), which also reflects on the very low percentage of outliers and a low NMAD value (see below).

Overall, we find a bias of $9.9\times10^{-4}$, a standard deviation of $0.0305$ and a NMAD of $0.021$. The $\sigma_{68}$ (i.e. the range in which falls the $68\%$ of the residuals) is $0.022$, smaller than the standard deviation, as it has to be expected from a \textit{leptokurtic} distribution.
Moreover, our method leads to a very small fraction of outliers, i.e. less than $0.39\%$ and $3.30\%$ using the $\left| \Delta z \right| > 0.15$ and  $\left| \Delta z \right| > 2\sigma$ criteria,
respectively. If we refer to the sample of objects with $IMAFLAGS\_ISO = 0$, the bias, standard deviation and NMAD become $0.00072$, $0.0288$ and $0.0207$, respectively. While the fraction of outliers is of $0.32\%$ and $3.26\%$. Thus, although the present approach is quite immune to systematic effects in photometry, we find a small improvement in the statistics when the sources in the masked regions are removed from the analysis.

\section{The photometric catalogue}\label{catalog}

To produce the final \zphot\, catalogue, we initially considered the multi-source KiDS catalogue, i.e. sources detected in $r$-band and measured in all KiDS bands. However, it is important to underline that all empirical \zphot\, prediction methods suffer from a poor capability to extrapolate outside the range of distributions imposed by the training sample. In literature, several approaches have been proposed to extend the applicability test of empirical methods outside the boundaries of the parameter space properly sampled by the spectroscopic KB (c.f. \citealt{vanzella2004,hoyle2015}). While useful in some cases, this artificial augmentation of the KB introduces a further level of complexity and leads to statistical biases which are difficult to evaluate and control.

In the available spectroscopic KB we found that $\sim99\%$ of the KB objects falls within the following region of the parameter space:
\begin{itemize}
 \item MAGAP\_4\_u $\leq25.1 $
 \item MAGAP\_6\_u $\leq24.7 $
 \item MAGAP\_4\_g $\leq24.5 $
 \item MAGAP\_6\_g $\leq24.0 $
 \item MAGAP\_4\_r $\leq22.2 $
 \item MAGAP\_6\_r $\leq22.0 $
 \item MAGAP\_4\_i $\leq21.5 $
 \item MAGAP\_6\_i $\leq21.0 $
\end{itemize}

Hence, to produce the final \zphot\, catalogue, we have removed all the objects that do not match the above criteria in more than one band.
The choice to retain objects with only one band not matching the above criteria was dictated by the need to maximize the number of objects with a redshift estimate and supported by the well tested robustness of the MLPQNA method against non detections or missing data \citep{cavuoti0}. In Table~\ref{TAB:stat} we report the statistical indicators evaluated for two groups of objects: those having all data points falling within the above region (\textit{clean} objects) and those (\textit{contaminated} objects) with only one band falling outside of it. It appears evident that for a one-band failure there is only a small decrease of performance.

\begin{table}
\resizebox{8.5cm}{!}{
\centering
\begin{tabular}{cccccc}
\hline
\hline
Subset             & $|bias|$  & $\sigma$   & NMAD    & Outliers $\%$              & Outliers $\%$                \\
                   &           &            &         & $|\Delta z|>0.15$   & $|\Delta z|>2\sigma$  \\
\hline
\hline
clean              & 0.0011   & 0.0303      & 0.0212  & 0.38                       & 3.13 \\
\hline
contaminated       & 0.0003   & 0.0339      & 0.0223  & 0.47                       & 5.80 \\
\hline
\end{tabular}}
\caption{Statistical indicators computed for two different subsets of the blind test set. The \textit{clean} set includes only data for which the photometry falls within the limits listed in Sec.~\ref{catalog}. The \textit{contaminated} subset includes the objects which fall outside those limits in only one band.} \label{TAB:stat}
\end{table}

However, in order to keep track of this effect, we include a \zphot\, quality flag in the catalogue, set to $1$ for \textit{best quality} (i.e. {\it clean}) and $0$ for  \textit{worse quality} (i.e. {\it contaminated}) objects.

The final \zphot\, catalogue consists of $1,142,992$ objects ($699,155$ objects have all $IMAFLAGS\_ISO = 0$ and $710,127$ with best quality).

\section{Conclusions}\label{conclusions}

In this work we applied the MLPQNA neural network to the ESO KiDS DR2 photometric galaxy data, using a knowledge base derived from the SDSS and GAMA spectroscopic samples, to produce a catalogue of photometric redshifts based on optical photometric data only. We obtained an overall $1\sigma$ uncertainty on $\Delta z = (z_{spec} - z_{phot}) / (1+ z_{spec})$ of $0.0305$ with a very small average bias of $9.9\times10^{-4}$, a low $NMAD$ of $0.021$, and a low fraction of outliers ($0.39\%$ above the standard limit of $0.15$).

The trained network was then used to process all galaxies in the data set that populate a parameter space similar to that defined by the SDSS+GAMA spectroscopic sample, producing  \zphot\, estimates for about $1.1$ million KiDS galaxies. The catalogue will be made available on CDS VizieR facility.

Deriving photometric redshifts is an essential task when dealing with large samples of galaxies, such as that expected from the KiDS photometric survey. These redshifts are currently being used by the Kids collaboration for a variety of studies regarding the evolution of galaxy stellar masses, integrated colours, colour gradients and the structural parameters with redshift (Napolitano et al. in prep.). The characterization of how completeness and biases of the photo-z catalogue affect the final scientific goals is therefore postponed to later works. This types of studies will allow us to better constrain the processes leading to the (mass) growth of galaxies in the last half of the current age of the universe.

\section{acknowledgements}
The authors would like to thank the anonymous referee for extremely valuable comments and suggestions.
Based on data products from observations made with ESO Telescopes at the La Silla Paranal Observatory under programme IDs 177.A-3016, 177.A-3017 and 177.A-3018, and on data products produced by Target/OmegaCEN, INAF-OACN, INAF-OAPD and the KiDS production team, on behalf of the KiDS consortium. OmegaCEN and the KiDS production team acknowledge support by NOVA and NWO-M grants. Members of INAF-OAPD and INAF-OACN also acknowledge the support from the Department of Physics \& Astronomy of the University of Padova, and of the Department of Physics of Univ. Federico II (Naples).
The authors would like to thank Amata Mercurio, Joachim Harnois-Déraps and Peter Schneider for the very useful comments.
CT has received funding from the European Union Seventh Framework Programme (FP7/2007-2013) under grant agreement n. 267251 \textit{Astronomy Fellowships in Italy} (AstroFIt).
This work was partially funded by the MIUR PRIN \textit{Cosmology with Euclid}.
MB acknowledges the support by the PRIN-INAF 2014 \textit{Glittering kaleidoscopes in the sky: the multifaceted nature and role of Galaxy Clusters}.


\phantomsection
\label{lastpage}
\end{document}